\documentclass[]{aa}
\usepackage[colorlinks=true,linkcolor=blue,citecolor=blue]{hyperref}
\usepackage[dvipsnames]{xcolor}
\usepackage{graphicx}
\graphicspath{{graphics}}
\usepackage{txfonts}
\usepackage[separate-uncertainty=true, list-units=repeat]{siunitx}
\usepackage{subcaption}
\usepackage{xspace}
\usepackage{hyperref}
\usepackage{stfloats}
\usepackage{svg}
\usepackage[inline]{enumitem}
\usepackage{physics} 	
\usepackage{multirow}
\usepackage{bm}
\usepackage{booktabs}
\usepackage[flushleft]{threeparttable}
\usepackage{makecell}

\bibpunct{(}{)}{;}{a}{}{,} 

\newcommand{\orcid}[1]{\href{https://orcid.org/#1}{\includegraphics[width=10pt]{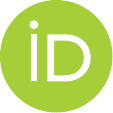}}}

\newcolumntype{M}[1]{>{\centering\arraybackslash}m{#1}}

\newcommand{\sextractor}{\textsc{SExtractor}\xspace}
\newcommand{\sepp}{\textsc{SourceXtractor++}\xspace}
\newcommand{\EAZY}{\textsc{eazy}\xspace}
\newcommand{\psfex}{\textsc{PSFEx}\xspace}

\newcommand{\hst}{\textit{HST}\xspace}
\newcommand{\JWST}{\textit{JWST}\xspace}
\newcommand{\NIRCAM}{NIRCam\xspace}

\DeclareSIUnit{\Msun}{M_\odot}
\DeclareSIUnit{\year}{yr}
\DeclareSIUnit{\pc}{pc}
\DeclareSIUnit{\mag}{mag}
\DeclareSIUnit{\mas}{mas}
\DeclareSIUnit{\dex}{dex}
\DeclareSIUnit{\jansky}{Jy}
\DeclareSIUnit{\kpc}{kpc}
\DeclareSIUnit{\arcsecond}{''}

\newcommand{\zphot}{$z_{\rm phot}$}

\begin{document} 

   \title{DAWN JWST Archive: Morphology from profile fitting of over 340,000 galaxies in major \JWST fields}
   \subtitle{Morphology evolution with redshift and galaxy type}  

   \titlerunning{Morphology from profile fitting of over 340,000 galaxies in major \JWST fields}
   \authorrunning{A. Genin et al.}

   \author{Aurélien~Genin\inst{\ref{DAWN},\ref{NBI},\ref{LX}}\thanks{Corresponding author. aurelien.genin@polytechnique.org}\orcid{0009-0005-0653-477X}
   \and
   Marko~Shuntov\inst{\ref{DAWN},\ref{NBI},\ref{UG}}\thanks{Corresponding author. marko.shuntov@nbi.ku.dk}\orcid{0000-0002-7087-0701} 
   \and
   Gabe~Brammer\inst{\ref{DAWN},\ref{NBI}}\orcid{0000-0003-2680-005X} 
   \and
   Natalie~Allen\inst{\ref{DAWN},\ref{NBI}}\orcid{0000-0001-9610-7950}
   \and
   Kei~Ito\inst{\ref{DAWN},\ref{DTU}}\orcid{0000-0002-9453-0381}
   \and
   Georgios Magdis\inst{\ref{DAWN},\ref{NBI},\ref{DTU}}\orcid{0000-0002-4872-2294}
   \and
   Jasleen~Matharu\inst{\ref{DAWN},\ref{NBI}}\orcid{0000-0002-7547-3385}
   \and
   Pascal~A.~Oesch\inst{\ref{DAWN},\ref{NBI},\ref{UG}}\orcid{0000-0001-5851-6649}
   \and
   Sune~Toft\inst{\ref{DAWN},\ref{NBI}}\orcid{0000-0003-3631-7176}
   \and 
   Francesco~Valentino\inst{\ref{DAWN},\ref{DTU}}\orcid{0000-0001-6477-4011}
    }

   \institute{
    Cosmic Dawn Center (DAWN), Denmark\label{DAWN}%
    \and%
    Niels Bohr Institute, University of Copenhagen, Jagtvej 128, 2200 Copenhagen, Denmark \label{NBI}%
    \and
    Ecole Polytechnique, Institut Polytechnique de Paris, Route de Saclay, 91120 Palaiseau, France \label{LX}
    \and%
    DTU Space, Technical University of Denmark, Elektrovej, Building 328, 2800, Kgs. Lyngby, Denmark\label{DTU}%
    \and
    University of Geneva, 24 rue du Général-Dufour, 1211 Genève 4, Switzerland\label{UG}
   }

   \date{Received ; accepted }

  \abstract
  {
Understanding how galaxies assemble their structure and evolve morphologically over cosmic time is a central goal of galaxy evolution studies. In particular, the morphological evolution of quiescent and star-forming galaxies provides key insights into the mechanisms that regulate star formation and quenching. We present a new catalog of morphological measurements for more than 340,000 sources spanning $0 < z < 12$, derived from deep \JWST NIRCam imaging across four major extragalactic fields (CEERS, PRIMER–UDS, PRIMER–COSMOS, GOODS) compiled in the DAWN \JWST Archive (DJA).
We performed two-dimensional surface brightness fitting for all galaxies in a uniform, flux-limited sample. Each galaxy was modeled with both a Sérsic profile and a two-component (bulge and disk) decomposition, yielding consistent structural parameters, including effective radius, Sérsic index ($n_S$), axis ratio, and bulge-to-total ratio ($B/T$).
To demonstrate the scientific application of our morphology catalogs, we combined these measurements with DJA photometric redshifts, physical parameters and rest-frame colors, and investigated the relation between total, bulge, and disk sizes, $n_S$ , star formation activity, and redshift. Bulge-dominated galaxies (high $n_S$ and $B/T$) predominantly occupy the quiescent region of the $UVJ$ diagram, while disk-dominated galaxies are mostly star-forming. A significant bimodality persists, with quiescent disks and compact, bulge-dominated star-forming galaxies observed out to $z > 3$. Quiescent galaxies also show significantly higher stellar mass surface densities, nearly an order of magnitude greater at $z \sim 4$ than at $z \sim 1$.
Our results confirm a strong and evolving link between morphology and star formation activity and support a scenario in which bulge growth and quenching are closely connected. This work is a highly valuable addition to the DJA, adding a morphological dimension to this rich dataset and thus enabling a wider scientific application.
  }


   \maketitle


\section{Introduction}

In the standard Lambda cold dark matter ($\Lambda$CDM) paradigm, galaxies form as gas accretes and cools within dark matter halos, which in turn shapes their sizes and morphologies \citep[e.g.,][]{FallEfsthathiou1980, Mo1998, Somerville2018}. Over their lifetimes, galaxies undergo various physical processes that regulate or disrupt their growth, such as star formation, feedback from stars and active galactic nuclei (AGNs), mergers, and gas accretion, all of which leave imprints on their structural properties. Consequently, studying galaxy morphology across cosmic time is key to understanding the physical transformations that accompany galaxy evolution, as well as the interplay between galaxies and their host dark matter halos \citep[e.g.,][for a review]{Conselice2014}.

Historically, galaxy morphology has been studied through a range of techniques that have evolved alongside observational capabilities. Early classifications followed the Hubble-de Vaucouleurs visual classification scheme \citep{1936rene.book.....H, 1959HDP....53..275D}. These visual schemes relied on manual inspection of photographic plates to identify structural features such as spiral arms, bars, and bulges, forming the basis for the first physical interpretations of galaxy evolution. With the advent of digital imaging, quantitative approaches became more common. Parametric methods, such as fitting analytic functions to galaxy light profiles, became standard, beginning with the de Vaucouleurs $r^{1/4}$ law for elliptical galaxies and generalized by the Sérsic profile \citep{1963BAAA....6...41S}, which remains widely used to characterize galaxy structure through parameters such as concentration, effective radius, and ellipticity. In parallel, nonparametric methods were developed to capture structural diversity without assuming specific functional forms, using parameters such as concentration, asymmetry, clumpiness (CAS), Gini coefficient, and $M_{20}$ to quantify galaxy morphology in a model-independent way \citep{2003ApJS..147....1C, 2004AJ....128..163L}.

A series of large surveys from both ground- and space-based observatories has systematically advanced our understanding of galaxy morphology across cosmic time. In the local Universe, the Sloan Digital Sky Survey (SDSS) enabled statistical studies of galaxy structure, revealing, for instance, that the size distribution at fixed luminosity follows a log-normal form at $z \lesssim 0.3$ \citep{2003MNRAS.343..978S}. The Galaxy Zoo project leveraged SDSS imaging to produce visual classifications for hundreds of thousands of galaxies \citep{2008MNRAS.389.1179L}, enabling studies of morphological diversity and its connection to environment and star formation \citep[e.g.,][]{schawinski_green_2014}. At intermediate redshifts, the Hubble Space Telescope (HST) played a transformative role through legacy programs such as GEMS \citep{2004ApJS..152..163R}, COSMOS \citep{2007ApJS..172..196K}, and CANDELS \citep{2011ApJS..197...35G, 2011ApJS..197...36K}, which enabled high-resolution imaging and morphological measurements out to $z \sim 3$. These datasets allowed for extensive morphological analyses, including evidence that quiescent galaxies are more compact, spheroidal, and concentrated than star-forming galaxies at fixed mass \citep[e.g.,][]{Daddi2005, Trujillo2007, Barro2013, Barro2017, 2016A&A...589A..35S}, and that they undergo size growth primarily via dry mergers \citep[e.g.,][]{2014ApJ...788...28V, VanDokkum2015, Whitaker2017}. However, the mechanisms that drive galaxy quenching remain debated, for example, whether morphological transformation and the cessation of star formation are causally linked, or whether they occur independently as consequences of other processes \citep[e.g.,][]{Tachella2015a}. Disentangling this connection requires large, statistically robust samples of galaxies with reliable morphological measurements spanning a wide redshift range, in order to trace structural evolution across cosmic time. Furthermore, at higher redshifts ($z \gtrsim 3$), HST imaging becomes limited to the rest-frame ultraviolet, which predominantly traces young stellar populations and can bias morphological interpretations \citep{2015ApJS..219...15S}. Nevertheless, these surveys laid the groundwork for the construction of large morphological catalogs \citep[e.g.,][]{2000AJ....120.1579Y, 2012ApJS..203...24V, Kawinwanichakij2021}, providing critical insights into the growth and structural transformation of galaxies over more than 10 billion years of cosmic history.

The James Webb Space Telescope (\JWST) has opened a new era in the study of galaxy morphology, enabling rest-frame optical measurements out to $z \gtrsim 7$ with unprecedented depth and resolution. Studies based on\JWST observations have revealed the evolution of galaxy morphology and size for both star-forming and quiescent galaxies out to unprecedented redshifts \citep[e.g.,][]{2024MNRAS.533.3724V, 2024MNRAS.527.6110O, Allen+2024, Matharu2024, Wright2024, Ito2024, yang2025cosmoswebunravelingevolutiongalaxy}. These studies show that quiescent galaxies remain structurally distinct from star-forming ones, with smaller effective radii and higher Sérsic indices, even at early times. Nevertheless, most \JWST morphological analyses to date have focused on specific galaxy selections, such as mass- or color-selected subsamples, or have been limited to individual fields. As a result, there is no consistent, flux-limited morphological catalog across different major \JWST fields and redshifts. This limits our ability to conduct comparative and statistical studies of galaxy structure, particularly for rarer populations such as quiescent galaxies at the full depth and breadth enabled by \JWST.

In this work, we build one of the largest and most uniform morphological catalogs to date using \JWST imaging, enabling consistent structural measurements across multiple fields and redshifts. We utilize publicly available mosaics from the DAWN \JWST Archive \citep[DJA,][]{Valentino2023}, which compiles deep NIRCam imaging from major extragalactic surveys, including CEERS, PRIMER (UDS and COSMOS), and GOODS. These are accompanied by matched photometric redshifts and physical parameters. To measure galaxy morphology, we carry out two-dimensional surface brightness profile fitting using \sepp \citep{2020ASPC..527..461B, kummel20}, a modern, scalable tool optimized for catalog-level model fitting in large multiband datasets. We fit each galaxy with both a single-component Sérsic profile \citep{1963BAAA....6...41S} and a two-component Bulge+Disk (B+D) model consisting of an exponential disk ($n_S=1$) and a de Vaucouleurs bulge ($n_S=4$). This dual-model approach allows us to probe structural diversity more flexibly and to derive key parameters, such as the Sérsic index ($n_S$), the effective radius, the axis ratio, and the bulge-to-total ratio ($B/T$). We apply this modeling to all sources above a flux and signal-to-noise threshold, yielding a homogeneous morphological catalog spanning approximately 340,000 sources across $0 < z < 12$. The aim of this catalog is to enable the community to carry out detailed statistical studies of galaxy morphology evolution throughout cosmic time.

This paper is organized as follows. The data and point spread function (PSF) reconstruction are presented in Section \ref{sec:data}. In Section \ref{sec:morphology}, we describe our method for measuring the morphology and sizes of galaxies across such a large field. In Section \ref{sec:results}, we present our results on the correlation between morphology and the $UVJ$ diagram, as well as the size evolution in the Sérsic and B+D models. We summarize our results in Section \ref{sec:conclusions}.
We adopt a \cite{2020A&A...641A...6P} $\Lambda$CDM cosmology with $H_0=67.4$\,km\,s$^{-1}$\,Mpc$^{-1}$, $\Omega_{\rm m,0}=0.315$. All magnitudes are expressed in the AB system \citep{oke_absolute_1974}.

\section{Data}\label{sec:data}

This work uses \NIRCAM images \citep{Rieke2005OverviewOJ} from public surveys of the \JWST, processed as part of the  \href{https://dawn-cph.github.io/dja/imaging/v7/}{DAWN JWST Archive (DJA)}, mosaic release v7. The DJA is an online repository containing reduced images, photometric catalogs, and spectroscopic data from public \JWST data, and is described in more detail in \cite{Valentino2023} and \cite{Heintz2025}.
In this work, we carry out morphological measurements in several major extragalactic surveys by \JWST, covering a total area of $\sim500$~arcmin$^2$. We focus on the following fields: 
\begin{enumerate}
    \item EGS from CEERS \citep[DD-ERS 1345][]{2023ApJ...946L..13F};
    \item GOODS from JADES \citep[GTO 1180, 1181, 1210, 1287][]{2023arXiv230602465E};
    \item UDS from PRIMER UDS \citep[GO 1837][]{2024MNRAS.tmp.1993D};
    \item COSMOS from PRIMER COSMOS \citep[GO 1837][]{2024MNRAS.tmp.1993D}.
\end{enumerate}
Table~\ref{tab:fields} presents the area and photometric bands used in these fields and their $5\sigma$ depths \citep[computed from empty apertures by][]{Weibel2024b}.

\subsection{Images}\label{sub:images}

\begin{table}[t!]
    \caption{\label{tab:fields}Summary of the fields covered in this work}
    \centering
    \begin{tabular}{M{1.4cm}ccM{3.6cm}}
        \multirow{2}{*}{Field} & Area & Depth & \multirow{2}{*}{Bands} \\
         & (arcmin$^2$) & (mag) & \\ \midrule \midrule
        EGS & 82.0 & 29.16 & F115W, F150W, F182M, F200W, F210M, F277W, F356W, F410M, F444W \\ \midrule
        GOODS & 67.3 & 29.93 & F090W, F115W, F150W, F200W, F277W, F356W, F444W \\ \midrule
        PRIMER-UDS & 224.4 & 28.51 & F090W, F115W, F150W, F200W, F277W, F356W, F444W \\ \midrule
        PRIMER-COSMOS & 127.1 & 28.62 & F090W, F115W, F150W, F200W, F277W, F356W, F444W \\ 
    \end{tabular}
    \tablefoot{All images are from the DJA and were processed prior to this work. The survey areas and $5\sigma$ depths computed from empty apertures correspond to the F277W band and were calculated by \cite{Weibel2024b}.}
\end{table}

The images used in this work are drawn from the DJA mosaic release v7. These are mosaics of the whole or split fields : \texttt{ceers-full-grizli-v7.2}, \texttt{gds-grizli-v7.2}, \texttt{gdn-grizli-v7.3}, \texttt{primer-uds-north-grizli-v7.2}, \texttt{primer-uds-south-grizli-v7.2}, \texttt{primer-cosmos-west-grizli-v7.0}, and \texttt{primer-cosmos-east-grizli-v7.0}.

The DJA provides reduced and calibrated images, in photometric units (10 nJy), which we used to run \sepp, alongside the weight maps that provide the inverse variance distribution. For source detection, we used the inverse-variance weighted stack of the long-wavelength filters (specified with \texttt{ir} in place of the filter in the image name). We did not apply any PSF convolution, since \sepp convolves the source models with the corresponding filter PSF (see Sect. \ref{sub:psf}).

\subsection{Catalogs}\label{sub:catalogs}

The DJA also provides photometric and photo-$z$ catalogs produced with \sextractor \citep{Bertin1996SExtractorSF} and \EAZY \citep{Brammer2008EAZYAF}. We used the DJA \sextractor catalogs as our principal list of sources for each field. Although we performed a separate detection with \sepp (Sect. \ref{sec:sepp}), we cross-matched with the DJA sources to ensure model-fitting-based measurements for the same sources. This approach also aligns with the DJA's provision of photo-$z$ and physical parameters from spectral energy distribution (SED) fittings with \EAZY on the \sextractor catalogs. In this work, we used the \EAZY output to investigate morphology evolution as a function of redshift and galaxy type (Sect. \ref{sec:results}).

\subsection{Point spread function reconstruction}\label{sub:psf}

Accurate model fitting requires precise characterization of the instrument's PSF. The PSF results from light diffraction at the aperture of the telescope and defines the resolution limit of \JWST. We empirically modeled the PSF from the final mosaics using \psfex \citep{2011ASPC..442..435B}, which builds a model for the PSF by fitting a set of basis functions to point sources provided in an input catalog.  Recent work by \cite{berman2024efficient} shows that \psfex accurately models the NIRCam PSF on mosaics.

\begin{figure}[t!]
    \centering
    \includegraphics[width=0.8\columnwidth]{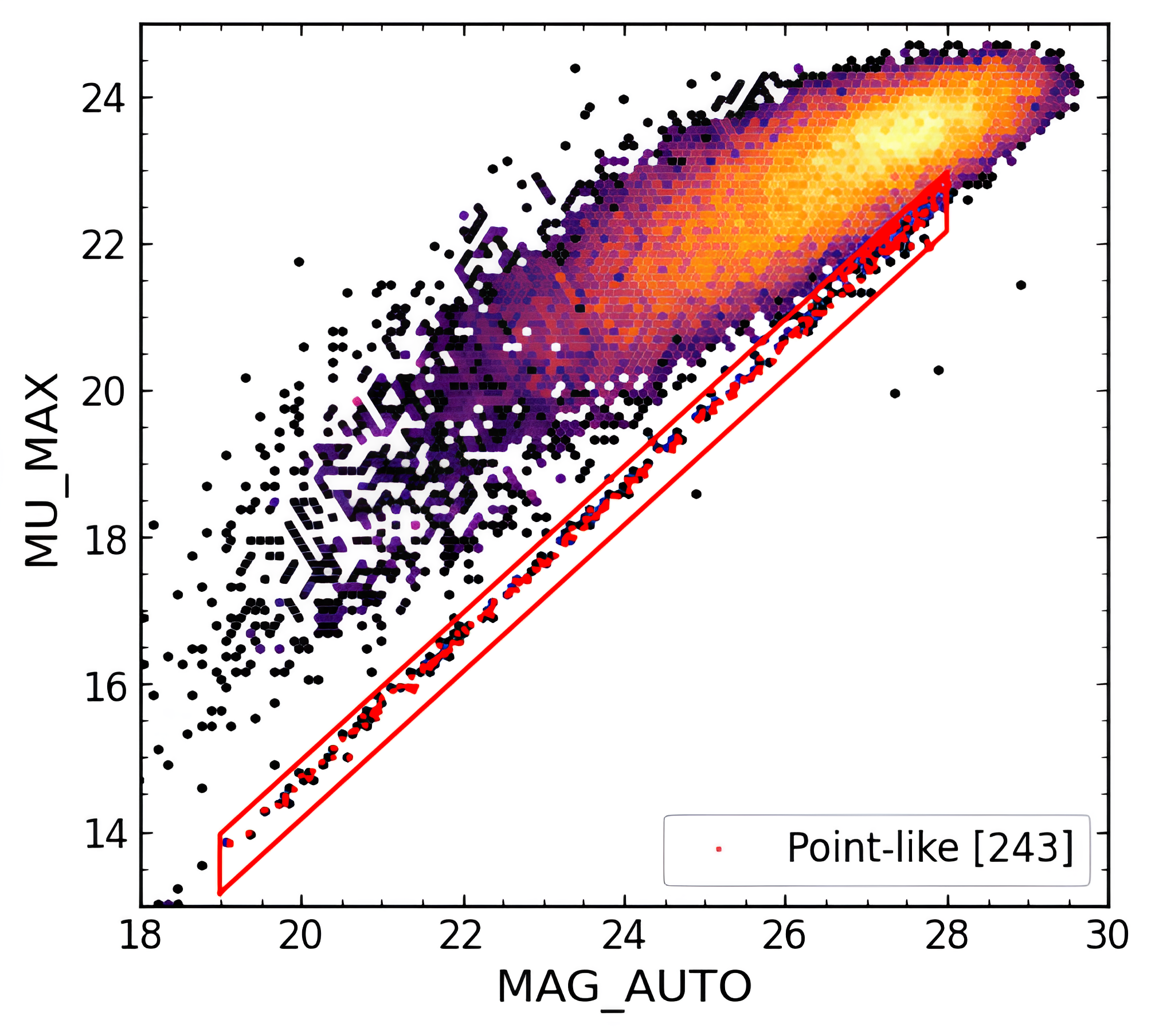}
    \caption{\footnotesize Point-like sources selection for PSF estimation. The plot shows the distribution (log scale) of sources from the GOODS-S field (F200W) in the \texttt{MU\_MAX/MAG\_AUTO} plane. The red box shows the starline selection of point-like sources, with the starline indicated by red dots.}
    \label{fig-psf:star-line}
\end{figure}

To select point sources, we developed a method inspired by \cite{2007ApJS..172..219L}. This method uses the \texttt{MU\_MAX} and \texttt{MAG\_AUTO} measurements from \sextractor. The parameter \texttt{MU\_MAX} represents the surface brightness of the brightest pixel of a source, while \texttt{MAG\_AUTO} denotes the Kron-like elliptical aperture magnitude. In the \texttt{MU\_MAX}/\texttt{MAG\_AUTO} plane (shown in Fig.~\ref{fig-psf:star-line}), the point-like sources follow a line of slope equal to one (referred to in this paper as the "starline"), and extended sources form a distinctive cloud above it. We added thresholds for \texttt{MAG\_AUTO}: a minimal value to avoid selecting saturated sources that would show a truncated PSF, and a maximal value to avoid including sources in the extended sources cloud.

Detection of the starline was performed using linear regression. However, to avoid bias from the extended sources cloud, we first applied a clustering algorithm to detect this point cloud and remove it. We used \texttt{DBSCAN} \citep{1996kddm.conf..226E}, implemented in the \textsc{scikit-learn} Python package. This algorithm identifies cores of high density (the extended sources cloud) while excluding points considered as noise (the starline). We then applied a threshold on the $\texttt{MU\_MAX}-\texttt{MAG\_AUTO}$ value to further remove outliers from the extended sources cloud. Finally, we detected the starline using the \texttt{RANSAC} linear regression algorithm \citep{fischler_bolles_1981}, implemented in \textsc{scikit-learn}. The threshold and width around the starline used to select point-like sources were chosen empirically on the basis of additional tests. Fig.\ref{fig-psf:star-line} shows the selection of point-like sources in the \texttt{MU\_MAX/MAG\_AUTO} plane.

The point-like source selection was expected to physically yield the same sample, independent of the band used. However, slight differences arose due to noise or differences in the threshold used for the $\texttt{MU\_MAX}-\texttt{MAG\_AUTO}$ value.
We chose the F200W band to select point-like sources, as it provided the most consistent starline detection across fields. Finally, we visually inspected a randomly selected sample ($N\sim100)$ to verify that the selected sources were point-like. Figure.\ref{fig-psf:psf} shows examples of the PSFs in the GOODS-S field.

\section{Morphological measurements}\label{sec:morphology}

\subsection{\sepp} \label{sec:sepp}

In this work, we present one of the largest catalogs of galaxy morphologies measured from \JWST images. To accomplish this, we used \sepp (hereafter abbreviated as SE++; \citep{2020ASPC..527..461B, kummel20}), the successor of the classic \sextractor \cite{Bertin1996SExtractorSF}, which performs brightness profile fitting to multiple sources in the multiband images. The principal advantage of SE++ is that it is optimized exactly for the purpose of large catalog creation -- it can simultaneously fit all sources in arbitrarily large mosaics and in different bands, without the need for prior sample selection, cutout creation, and masking. Additionally, its flexible model fitting allows the user to define any model using a simple Python configuration file. In the following, we describe our models and catalog-making procedure.

\subsection{Brightness profile models}\label{sub:models}

To measure galaxy morphology, we modeled the brightness profile of each source. The \sepp tool provides full control over the model. To maximize scientific applications, we used the two most common models:

\begin{enumerate}
    \item The Sérsic model \citep{1963BAAA....6...41S}, parameterized by a single shape parameter, the Sérsic index $n_S$; the effective radius $R_{e}$; the axis ratio $(a/b)$; the position angle $\theta$; and the total flux $f_{tot}$. In this work, we fit for the two components of the ellipticity, $e_1$ and $e_2$, which are directly linked to $(a/b)$ and $\theta$. The priors used for these parameters are presented in Fig.\ref{fig:priors}.
    
    \item The Bulge + Disk (B+D) model, a composite of an exponential disk ($n_S=1$) and a de Vaucouleurs bulge ($n_S=4$) Sérsic profile. It is parametrized by the effective radii of the disk, $R_{disk}$, and bulge, $R_{bulge}$; the axis ratios $(a/b)_D$ and $(a/b)_B$; a common angle $\theta_{BD}$; the total flux of both components, $f_{BD,tot}$; and the bulge-to-total ratio $B/T = f_{B,tot}/f_{BD,tot}$. The $B/T$ prior is a bell curve ranging from $5\times 10^{-5}$ to $1$, with a mean and spread that increase as a function of wavelength.
\end{enumerate}

The B+D model generally provides a better fit than the Sérsic model, as it more accurately describes the central region of galaxies, particularly the presence of a bulge or a dimmer center compared to a Sérsic profile. However, because it has more parameters, it is more computationally intensive and can, in some cases, lead to degeneracy in the model parameters, especially for fainter sources where the two components may not be distinguishable.

The SE++ tool performs model fitting simultaneously on multiple images, typically across different bands. In this work, we constrain the morphological parameters ($R_{\rm eff}$, $n_S$, $(a/b)$, and $\theta$) to be identical across all bands for a given source. This approach yields final values that represent a weighted average over the entire wavelength range used ($0.8-5 \si{\um}$). The SE++ configuration file defining the models is available on GitHub\footnote{\url{https://github.com/AstroAure/DJA-SEpp/blob/main/config/sepp-config.py}}.

\subsection{Tiling}\label{sub:tiling}

It is theoretically possible to run \sepp directly on the full mosaics from the DJA. However, because of speed and computing power limitations, we chose to tile the full images. This approach makes sources modeling faster, thanks to parallelization. We divided the mosaics into $2\si{\arcmin} \times 2\si{\arcmin}$ tiles, with $0.5\si{\arcmin}$ overlap, to ensure that sources fragmented by one tile were fully present in the neighboring tile.

\subsubsection{Catalog merging}\label{sub:tiling-cat}

To merge the sub-catalogs, we iterated over each tile to append them and create a catalog covering the whole field. At each iteration step, we cross-matched with the previously appended sub-catalogs and discarded the matched duplicates from the overlap regions. For sources that are matched (in the overlap region), we chose the one with the smallest uncertainty in the F200W magnitude, as measured by \sepp. As the model fitting was performed independently for the Sérsic and B+D models, this step was performed separately to produce one catalog per model. The F200W band was chosen because it has better resolution compared to the LW channel, and galaxies are generally brighter in it than in other bands of the SW channel.

Finally, we cross-matched the two model-fitting catalogs with the DJA catalogs. This step allowed us to remove false detections (which were frequent near the edges of images), retain the same list of sources, and provide additional morphological measurements to the DJA catalogs. 
We chose a threshold of 0.3\si{\arcsecond} as the cutting distance to validate a match. This threshold was set manually using the histogram of angular distances produced by the cross-match. This value corresponded to 5 pixels in the LW channel and 10 pixels in the SW channel of \NIRCAM, which was acceptable and could be the result of differences between the source centroids estimated in \sextractor and \sepp.

\subsubsection{Merging of the model and residual images}\label{sub:tiling-img}

The SE++ tool produces model and residual images. We merged only the model sub-images and generated a mosaic residual image afterward. To merge the sub-images, we used the \texttt{reproject\_and\_coadd} function from the \texttt{reproject} Python package \citep{2018zndo...1162674R}, which reprojected and co-added the images on a frame specified by the World Coordinate System (WCS) of the native DJA images. This ensured that the merged full model images had the same pixel scale, center, and orientation as the DJA images. We used this to generate mosaic residual images by subtracting the mosaic model images from the source DJA images.

\subsection{Flagging and completeness}\label{sub:completeness}

\begin{table*}[t!]
    \caption{\label{tab:completeness}Completeness of the morphology measurements using \sepp}
    \centering
    \begin{tabular}{@{}crrrr@{}}
    \textbf{Field} & \multicolumn{1}{c}{\textbf{DJA}} & \multicolumn{1}{c}{\textbf{Sérsic}} & \multicolumn{1}{c}{\textbf{Bulge+Disk}} & \multicolumn{1}{c}{\textbf{Both}} \\ \midrule\midrule
    \textbf{CEERS} & 67,035 & 52,604 \small{(78.5\%)} & 59,046 \small{(88.1\%)} & \textbf{51,329 \small{(76.6\%)}} \\ \midrule
    \textbf{GOODS-S} & 57,355 & 44,931 \small{(78.3\%)} & 52,754 \small{(92.0\%)} & \textbf{44,016 \small{(76.7\%)}} \\ \midrule
    \textbf{GOODS-N} & 65,481 & 53,291 \small{(81.4\%)} & 58,852 \small{(89.9\%)} & \textbf{51,465 \small{(78.6\%)}} \\ \midrule
    \textbf{PRIMER-UDS (N)} & 68,857 & 58,947 \small{(85.6\%)} & 67,134 \small{(97.5\%)} & \textbf{57,945 \small{(84.2\%)}} \\ \midrule
    \textbf{PRIMER-UDS (S)} & 65,864 & 57,397 \small{(87.1\%)} & 64,537 \small{(98.0\%)} & \textbf{56,476 \small{(85.7\%)}} \\ \midrule
    \textbf{PRIMER-COSMOS (E)} & 50,655 & 42,359 \small{(83.6\%)} & 48,496 \small{(95.7\%)} & \textbf{41,597 \small{(82.1\%)}} \\\midrule
    \textbf{PRIMER-COSMOS (W)} & 51,362 & 40,493 \small{(78.8\%)} & 46,964 \small{(91.4\%)} & \textbf{39,704 \small{(77.3\%)}} \\ \midrule \midrule
    \textbf{Total} & 426,609 & 350,022 \small{(82.0\%)} & 397,783 \small{(93.2\%)} & \textbf{342,892 \small{(80.4\%)}}
    \end{tabular}
    \tablefoot{Values in the table show only sources with F277W magnitudes below the $5\sigma$ depth of each field, and a S/N > 3. For the \sepp columns, the values correspond to sources successfully fitted (\texttt{flag}=2); see Sect. \ref{sub:completeness} The percentages are relative to the DJA catalogs with the same magnitude and S/N cut.}
\end{table*}

\begin{figure}[t!]
    \centering
    \includegraphics[width=0.9\columnwidth]{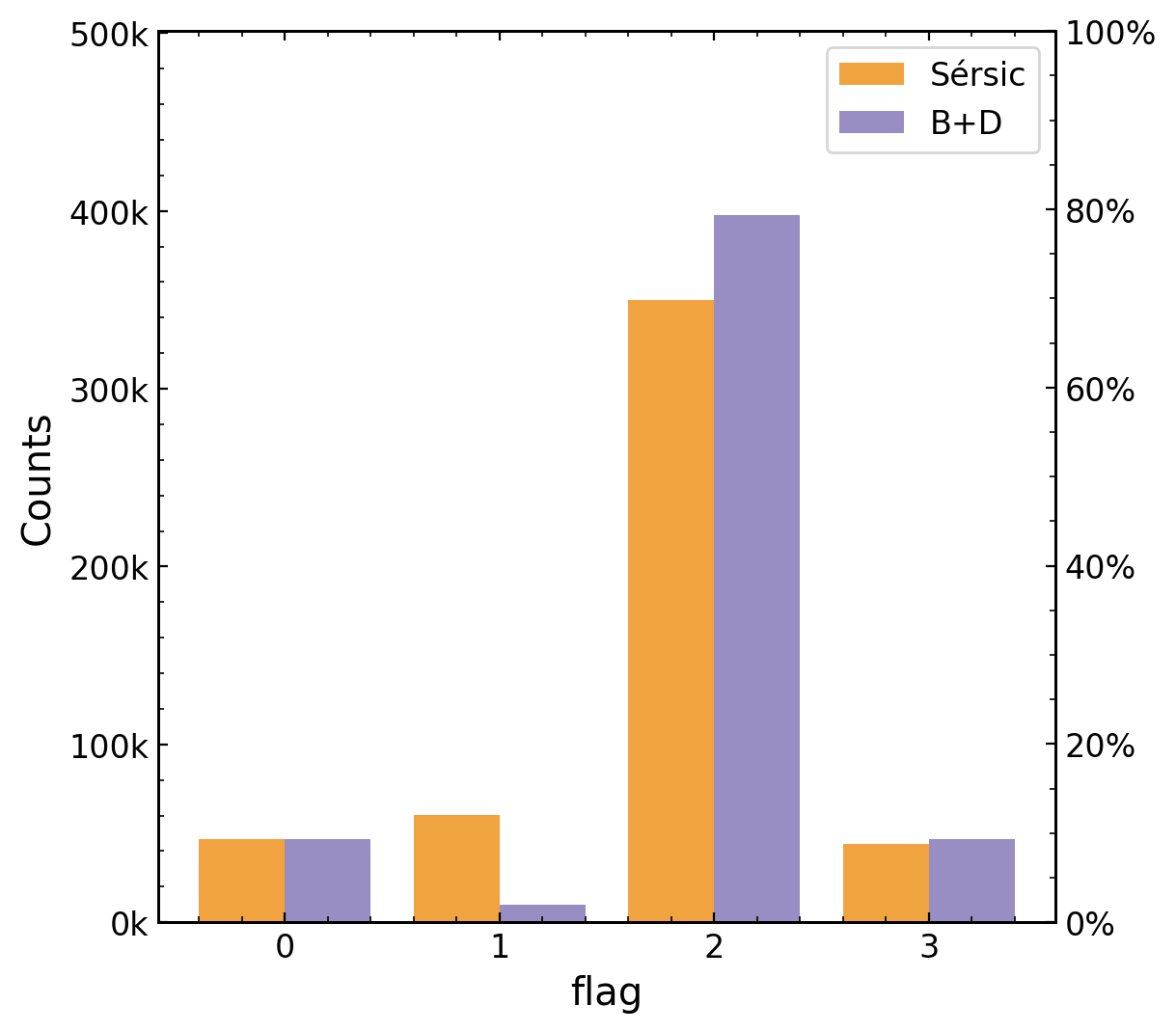}
    \caption{\footnotesize Histogram of the \texttt{flag} values for the full catalogs. The percentages are relative to the total number of sources in the DJA catalogs. Two distinctive histograms for the Sérsic and Bulge+Disk models are shown because their fitting success rates differ. The figure shows a vast majority (70-80\%) of sources are correctly fitted and provides science-ready morphological data.}
    \label{fig:flag}
\end{figure}

As mentioned previously, our aim was to add morphological measurements to the DJA catalogs. Therefore, by cross-matching, we retained the same number of rows (sources) in our catalogs as in the DJA catalogs. Furthermore, we added a \texttt{flag} keyword with four possible values as follows:
\begin{enumerate}
    \item[\textbf{0}]: The source was not fitted (no morphological data).
    \item[\textbf{1}]: A potential artifact occurred during model fitting.
    \item[\textbf{2}]: Fitting was performed successfully. This value gives science-ready data.
    \item[\textbf{3}]: The source has S/N < 3 or a magnitude fainter than the 5$\sigma$ depth of its corresponding survey. The S/N was computed by \sepp on the detection image.
\end{enumerate}
The distribution of the \texttt{flag} values is shown in Fig.\ref{fig:flag}. Seventy to eighty percent of the sources were successfully fitted (\texttt{flag=2}). This indicates that the \sepp minimization algorithm converged and the parameter values were not equal to the initial values nor at the minimum or maximum of the allowed range. Sources without morphological data (\texttt{flag=0}) were not detected in the \sepp detection step and are likely artifacts or faint sources near the image depth, which would be difficult to fit nonetheless. Potential artifacts (\texttt{flag}=1) are sources that have been fitted, but with some parameters remaining at their initial values or that strayed to the extreme ends of the allowed parameter range. Typically, this indicates a poor fit; therefore, we flagged such sources. For the Sérsic model, this includes sources with $(a/b) > 0.99$, $\abs{n_S-0.36}<10^{-4}$, $n_S>8.35$, or $n_S<0.301$. For the B+D model, this includes $(a/b)_D>0.9999$, $(a/b)_D<0.10001$, or $\abs{(a/b)_D-0.5}<10^{-5}$. These values were chosen manually by identifying artifacts in the parameter distribution and by visual inspection. These sources should be handled with care by the user.

Since the list of detected and measured sources by \sepp is slightly different from the initial DJA list, we quantified the completeness of our morphological catalogs. Table~\ref{tab:completeness} presents the number of galaxies in each field used in the DJA, and the number of successfully fitted galaxies (\texttt{flag}=2) in each model (Sérsic and B+D) from our work. 

The completeness was measured relative to the number of galaxies in the DJA catalogs, using the same quality cut on S/N and magnitude as S/N $> 3$ and mag $< 5\sigma$ depth (\texttt{flag}=3) on both. In total, our catalog consists of 342,892 sources with reliable model fitting and science-ready, making it one of the largest morphological catalogs based on \JWST observations.

A total of 80.4\% of the DJA S/N $> 3$ and mag $< 5\sigma$ depth sources were successfully fitted with both a Sérsic and a B+D model. The completeness is higher for the B+D model than for the Sérsic model. This is likely because the B+D model is a better description for some sources whose Sérsic model parameters tend toward the minimum or maximum allowed values (therefore classified as \texttt{flag}=1). Finally, we analyzed the distribution of certain physical parameters (F277W magnitude, \zphot, mass, and Kron radius) for the non-detected sources (\texttt{flag}=0) and potential artifacts (\texttt{flag}=1). We did not find any significant correlation between these parameters, indicating that the incompleteness due to \texttt{flag}=1 is not biased with magnitude, redshift or stellar mass. 

Therefore, we consider our morphological catalogs to have a relatively high completeness compared to the DJA, and, as such, they provide highly valuable information for studying the morphology of galaxies and its evolution through cosmic time, as demonstrated in Sect. \ref{sec:results}.

\subsection{Comparisons with previous work}\label{sub:validation}

To validate our measurements, we compared them with previous morphological catalogs in the same fields, particularly \cite{2012ApJS..203...24V}. These catalogs contain Sérsic modeling performed using \textit{Hubble Space Telescope} (\hst) observations in the same fields used in this work: GOODS, EGS, UDS, COSMOS. By cross-matching them with our catalog and selecting only early-type galaxies (see Eq.\ref{eq:quiescent}), we find 3,263 matches with S/N $>10$.

The size and morphology of galaxies are known to differ in different wavelength ranges. To ensure that we compared morphologies measured at similar observer-frame wavelengths, we used the measurements in F160W from \cite{2012ApJS..203...24V}.
However, since our measurements correspond to the averaged morphology over NIRCam's wavelength range ($\sim1-5 \si{\um}$), we scaled the effective radii $R_{\rm eff}$ in \cite{2012ApJS..203...24V} to $2.5 \si{\um}$ using the following scaling relation:
\begin{align}
    \label{eq:correct-radii}
    \frac{\Delta \log R_{\rm eff}}{\Delta \log \lambda} = -0.25,
\end{align}
which was found by \cite{2012ApJS..203...24V} to describe the wavelength dependence of $R_{\rm eff}$ for early-type galaxies. We note that a more complex version of this relation has been proposed by the same authors, which integrates the effect of redshift and stellar mass. However, we adopted this simpler and more general relation for comparison purposes.

\begin{figure}[h!]
    \centering
    \includegraphics[width=\columnwidth]{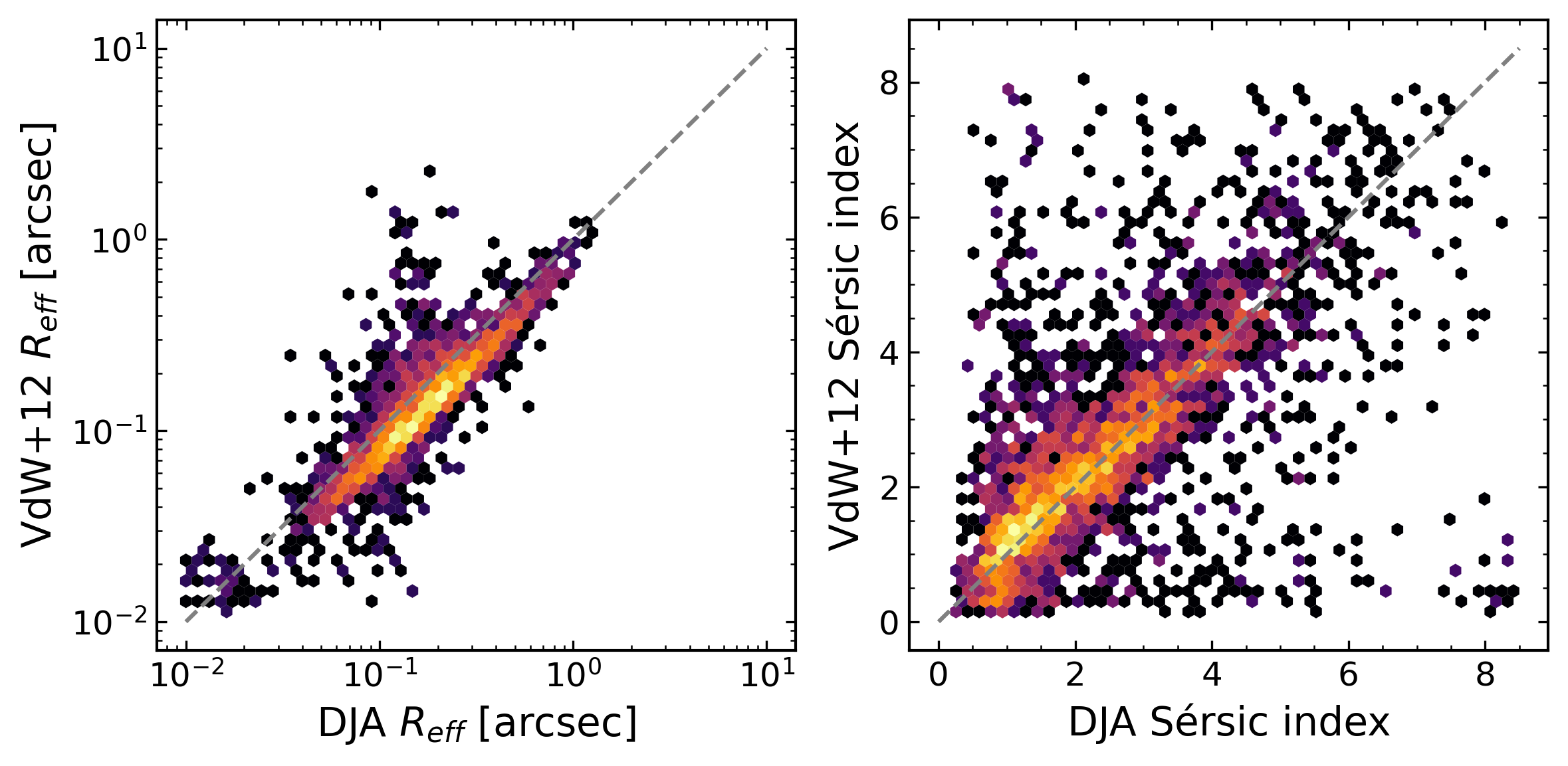}
    \caption{\footnotesize Comparison between Sérsic model fitting in our work and \cite{2012ApJS..203...24V} for UVJ-selected quiescent galaxies. The figure compares effective radii $R_{\rm eff}$ and Sérsic index $n$. We corrected the $R_{\rm eff}$ values from \cite{2012ApJS..203...24V}, originally measured at $1.6 \si{\um}$, to $2.5 \si{\um}$ using Eq.\ref{eq:correct-radii},  in order to match the average wavelength of our \JWST images. The color of the hexbin cells indicates the counts in each cell in log scale. Gray dotted lines show identical values between the two works. This figure includes 3,263 matched galaxies.}
    \label{fig:hst-comparisons}
\end{figure}

Figure~\ref{fig:hst-comparisons} shows the one-to-one comparisons between the two works for $R_{\rm eff}$ (top middle panel) and the Sérsic index $n_{\rm S}$ (top right panel), as well as the histograms of $R_{\rm eff}$, $n_{\rm S}$, and the axis ratio ($a/b)$ (bottom panels). In general, there is good agreement in the effective radii, with a slight trend toward larger radii from our measurements. One reason for the slight offset could be the scaling using Eq.~\ref{eq:correct-radii} that we applied. Without this scaling, the trend inverses and our effective radii are smaller than those of \cite{2012ApJS..203...24V}. 

\section{Morphological evolution with redshift and galaxy type}\label{sec:results}

To demonstrate the scientific application of our morphological catalogs, we investigated statistical distributions of several morphology indicators as a function of redshift and galaxy type. Our measurements cover a wide range of redshifts, enabling studies of morphological evolution across cosmic time. For this analysis, we used the photo-$z$, rest-frame colors, and other physical parameters from the DJA \EAZY catalogs. In particular, we investigated morphological evolution as a function of galaxy type (quiescent vs. star-forming).

\subsection{$UVJ$ diagram as a function of morphology and redshift}\label{sub:morpho-uvj}

To investigate the relationship between galaxy morphology, type, and redshift, we used the $UVJ$ color diagram to select star-forming and quiescent galaxies \citep{Williams2009}. In principle, the $UVJ$ selection criteria for quiescent galaxies change with redshift \citep{Williams2009}. However, for our qualitative analysis, we adopted a redshift-independent selection corresponding to the $0.7<z<1.3$ range used in \cite{2016A&A...589A..35S}, who also investigated the morphology dependence of the $UVJ$ diagram from \hst observations:
\begin{equation}\label{eq:quiescent}
UVJ_{\text{quiescent}} = 
    \begin{cases}
        U-V > 1.3 \text{ , and} \\
        V-J < 1.6 \text{ , and} \\
        U-V > 0.88(V-J) + 0.49 \text{.}
    \end{cases}
\end{equation}
In addition, we focused our analysis on $\log{M_\star/\si{\Msun}} > 10$, \texttt{flag}=2, and S/N $>10$, resulting in 13,685 galaxies. This ensured that our sample had sufficiently high S/N and robust morphological estimates from the model fitting.

\begin{figure*}[t!]
    \centering
    \includegraphics[width=\textwidth]{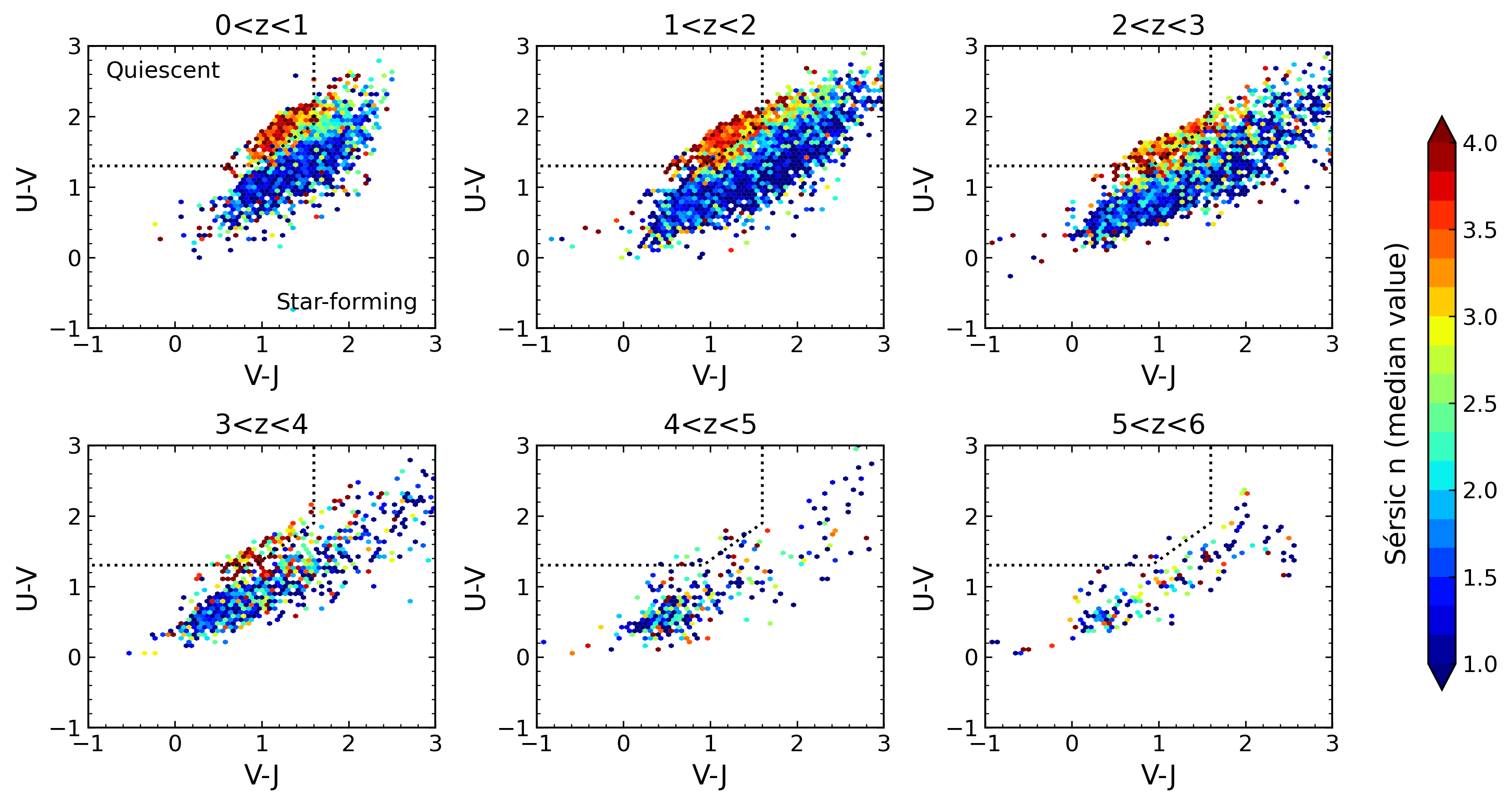}
    \caption{Distribution of Sérsic indices, $n_S$, in the $UVJ$ color space for different redshift ranges, for galaxies with $\log{M_\star/\si{\Msun}} > 10$. The color of each hexagonal bin represents the median value of $n_S$, as indicated by the colorbar. The dotted line shows the quiescent vs star-forming separation using Eq.~\ref{eq:quiescent}. The quiescent region is predominantly populated by galaxies with high Sérsic indices ($n_S\gtrsim 3$).}
    \label{fig-uvj:sersic}
\end{figure*}

First, we investigate the location of galaxies in the $UVJ$ diagram as a function of the Sérsic index, $n_S$, and redshift. Fig.~\ref{fig-uvj:sersic} shows the $UVJ$ diagram in six redshift bins at $0<z<6$ (in different panels), color coded by the Sérsic index. The dotted lines mark the regions separating star-forming and quiescent galaxies (Eq.~\ref{eq:quiescent}).
Fig.~\ref{fig-uvj:sersic} shows a clear correlation between the Sérsic index and $UVJ$ colors. There is a gradient of the Sérsic index, $n_S$, in a direction roughly orthogonal to the boundary between quiescent and star-forming galaxies, such that galaxies with a higher Sérsic index preferentially populate the redder and quiescent $UVJ$ region. This trend holds for the different redshift bins. For $z>4$, the samples are too small to draw statistically robust conclusions, especially for quiescent galaxies.

\begin{figure*}[t!]
    \centering
    \includegraphics[width=0.8\textwidth]{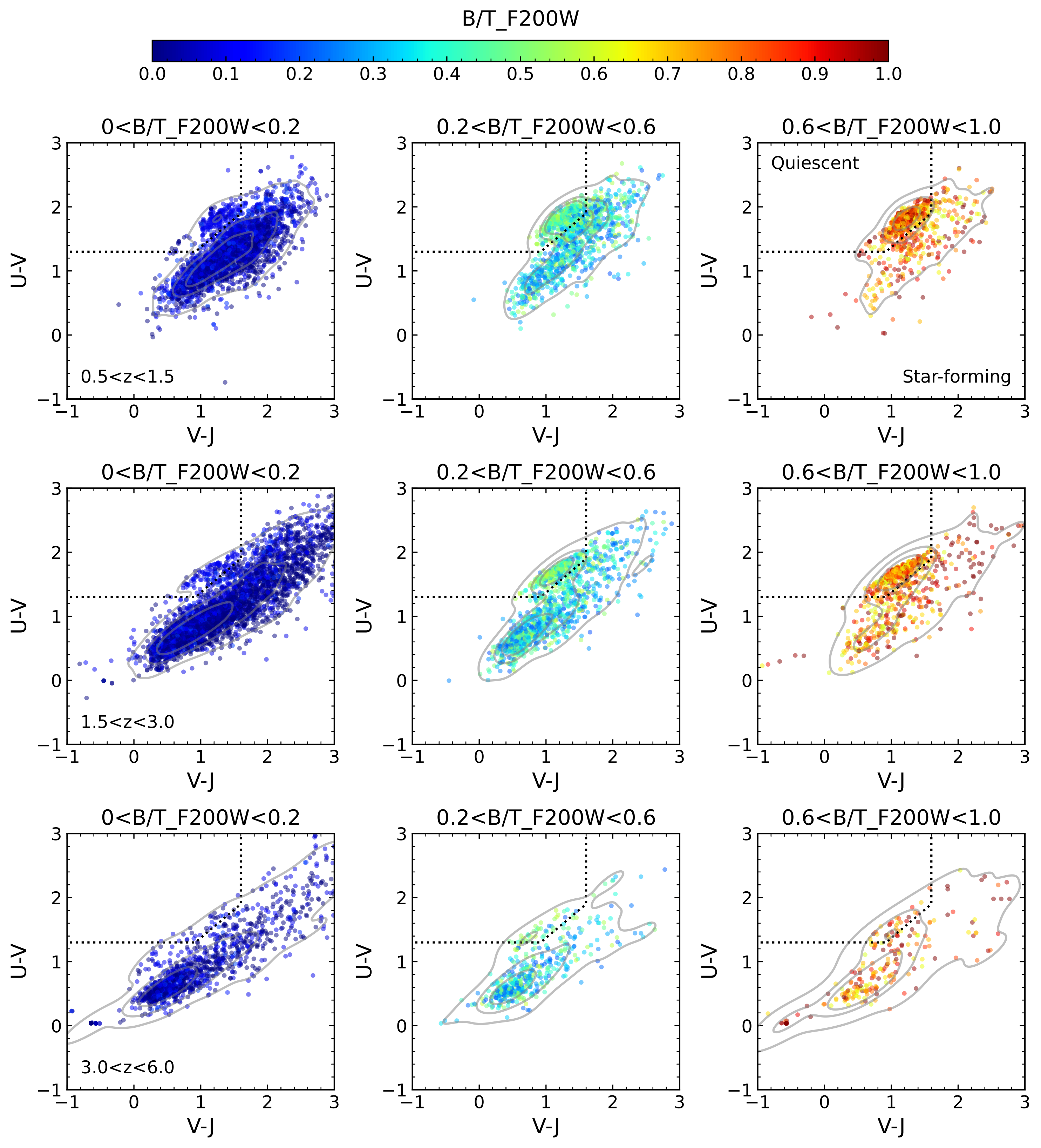}
    \caption{Distribution of  $B/T$ in the $UVJ$ color space for different redshift and $B/T$ ranges, for galaxies with $\log{M_\star/\si{\Msun}} > 10$. Each point is colored by its $B/T$ value in the F200W band. The dotted line shows the quiescent versus star-forming separation using Eq.~\ref{eq:quiescent}. The contours indicate 25, 50, and 75\% of the density. Bulge-dominated galaxies (high $B/T$) predominantly occupy the quiescent region, whereas disk-dominated galaxies (low $B/T$) are generally in the star-forming region of the $UVJ$ space.}
    \label{fig-uvj:BT}
\end{figure*}

Second, because we also used B+D models to fit the galaxies in our catalogs, we could analyze how their $B/T$ values populate the $UVJ$ diagram. For this analysis, we used the $B/T$  measured in the F200W band because it offers better resolution, being in the SW channel. Figure\ref{fig-uvj:BT} shows $UVJ$ diagrams for three broad redshift ranges and three $B/T$ ranges. The color indicates the $B/T$ value, with the color scale indicated at the top. We also show the contours estimated by kernel density. Fig.~\ref{fig-uvj:BT} shows that bulge-dominated galaxies ($B/T>0.6$) preferentially occupy the quiescent region, whereas disk-dominated galaxies ($B/T<0.2$) occupy the star-forming region. Intermediate galaxies that show both bulges and disks ($0.2<B/T<0.6$) form a diverse population that can be classified as both star-forming and quiescent based on their $UVJ$ colors. However, there is a trend showing that galaxies with a higher $B/T$ preferentially occupy the quiescent $UVJ$ region. At $z>3$ there are more galaxies with higher $B/T$ that are star-forming than quiescent. This results from the relative rarity of quiescent galaxies at these epochs, as well as from the presence of a population of compact star-forming galaxies \citep[e.g., blue nuggets,][which we discuss further in the next part of this section]{Barro2013, Dekel2014}. These qualitative trends from the independently fitted B+D model are in good agreement with those from the Sérsic model. This consistency attests to the robustness of the morphological measurements.

 \begin{figure*}[t!]
    \centering
    \includegraphics[width=0.9\textwidth]{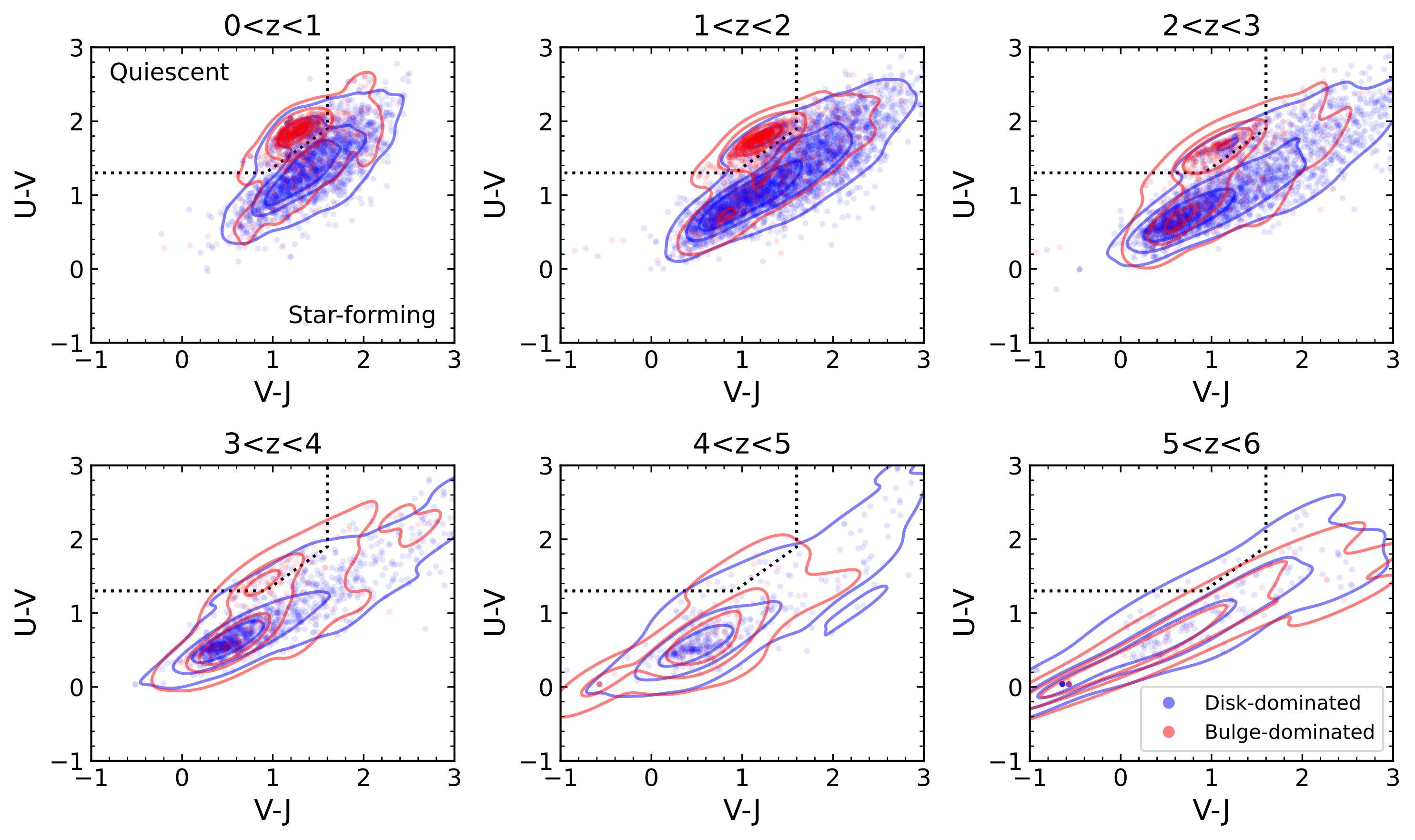}
    \caption{$UVJ$ color diagram for bulge- and disk-dominated $\log{M_\star/\si{\Msun}} > 10$ galaxies. The classification combines criteria based on both  $n_S$ and  $B/T$, estimated independently. Contour lines show kernel density estimates corresponding to 25, 50, and 75\% of the density.}
    \label{fig-uvj:classification}
\end{figure*}

To further investigate the correlation between bulge- and disk-dominated morphology and star formation activity, we defined two classes based on the two independent morphology estimates, $n_S$ and $B/T$, as follows:
\begin{enumerate}
    \item {Bulge-dominated galaxies}: $n_S>1$ and $B/T>0.5$.
    \item {Disk-dominated galaxies}: $n_S<4$ and $B/T<0.5$.
    \item 
\end{enumerate}
With this, we attempted to provide a robust binary classification of whether a galaxy is disk- or bulge-dominated and explore how they populate the $UVJ$ color space.

In Fig.\ref{fig-uvj:classification}, we show the $UVJ$ diagrams in the same redshift ranges as in Fig.\ref{fig-uvj:sersic}, with color indicating the two different classes: red for bulge-dominated galaxies and blue for disk-dominated galaxies. We plotted kernel density contour lines for each class. This classification is consistent with the broad conclusion that quiescent galaxies are bulge-dominated and star-forming galaxies are disk-dominated.

The disk-dominated sample still exhibits a degree of bimodality, with a small population of quiescent galaxies. Indeed, while star formation is typically associated with disk galaxies, quiescent disks do exist. At lower masses, quenching is often environmentally driven, particularly via strangulation \citep{Larson1980, Moran2007}, ram pressure stripping \citep{Gunn1972}, and galaxy harassment \citep{Moore1996, Moore1998} in dense environments \citep{peng_mass_2010, Cortese2021}. At higher masses, some fast-rotating quiescent disks can form through a combination of mild dissipative contraction and secular evolution \citep{Toft2017, DEugenio2024}. These quiescent disks may also result from gas exhaustion following a compaction event or mergers that preserve disk kinematics but suppress star formation \citep{Omand2014, Zolotov2015}.

Bulge-dominated galaxies preferentially occupy the quiescent $UVJ$ region out to $z\sim3$, with a small bimodality appearing in the star-forming region at higher redshifts. This indicates that bulges can still be actively forming stars, especially at earlier cosmic times, but by $z < 1$, most have migrated toward quiescence.  This trend may indicate a quenching pathway, in which bulge growth coincides with or precedes quenching, consistent with theoretical and observational evidence for compaction-driven quenching \citep[e.g.,][]{Barro2013, Barro2014, Tachella2015a, Tacchella2016}. The location of bulge-dominated galaxies in the $UVJ$ diagram -- toward bluer colors compared to the main disk-dominated population -- supports a fast quenching scenario, in which lower-mass, young, star-forming galaxies undergo rapid quenching \citep[$t_{\rm Q}<2$ Gyr e.g.,][]{Moutard2018, Belli2019}. This is consistent with the blue nugget phase, where galaxies experience a burst of compact star formation before quenching into red nuggets \citep[e.g.,][]{Dekel2014, Zolotov2015, Barro2017}. 
Finally, the lower number of bulge-dominated galaxies in the higher redshift ranges is consistent with the inverted Hubble classification evolution: galaxies form as disks and evolve to bulges \citep[e.g.,][]{2022A&A...666A.170Q}.

This qualitative analysis of the correlation between quenching and morphology is consistent with the current picture from both theory and observations. Importantly, by providing morphological measurements from \JWST for such a large sample, our work paves the way for more in-depth and quantitative population studies that can unveil the details of galaxy quenching and the accompanying morphological transformations.

\subsection{Size evolution}\label{sub:size-evolution}

\begin{figure*}[t!]
    \centering
    \includegraphics[width=\textwidth]{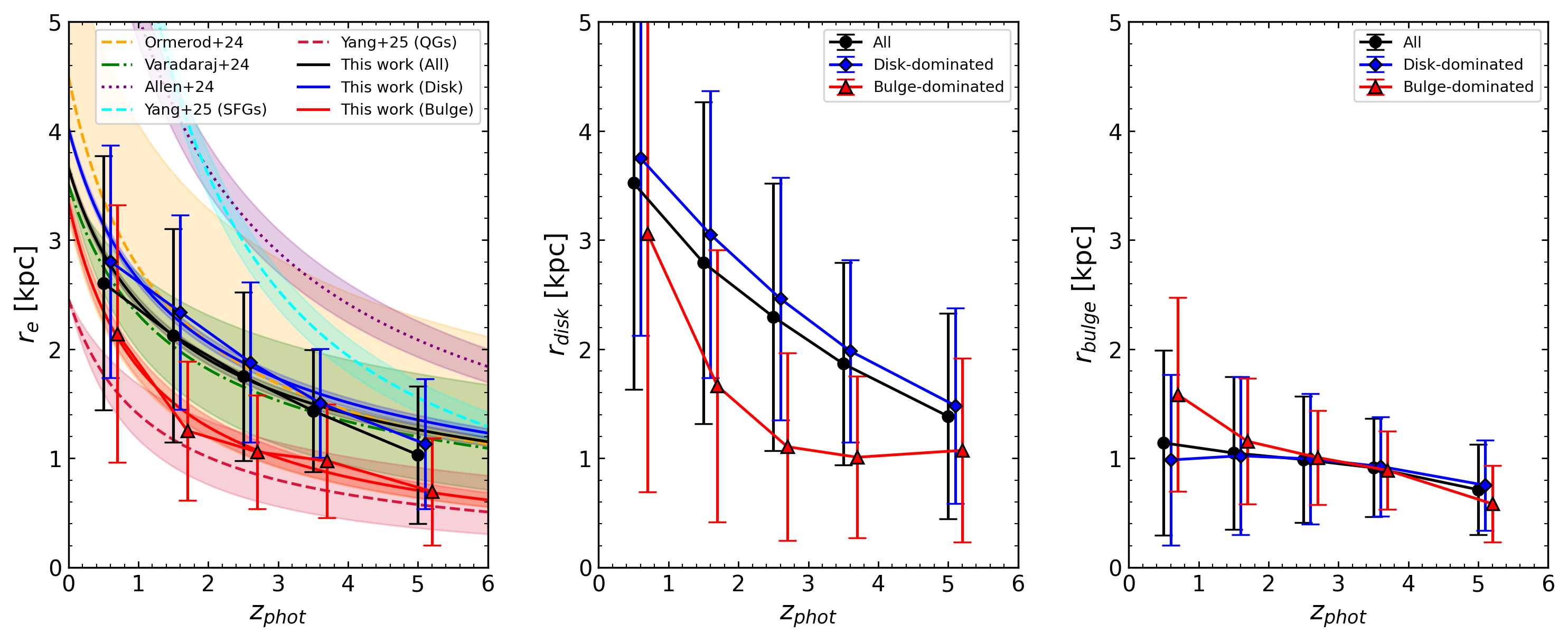}
    \caption{Evolution of galaxy sizes as a function of redshift. The plots show the Sérsic effective radius $r_e$ (left), the disk effective radius $r_{disk}$ (center), and the bulge effective radius $r_{bulge}$ (right) as a function of \zphot\ for $\log{M_\star/\si{\Msun}} > 10$ galaxies. This corresponds to a mean mass of $\approx4\times10^{10} \, \si{\Msun}.$  The points with error bars show the mean and $1\,\sigma$ at a given redshift for all galaxies (black), disk-dominated galaxies (blue), and bulge-dominated galaxies (red). The $r_e$ evolution is modeled in the left panel for all galaxies, as well as for the bulge- and disk-dominated populations, following $r_e=R(1+z)^a$ (see Sect. \ref{sub:size-evolution}). We compare the size evolution for all galaxies with previous \JWST-based work from \cite{2024MNRAS.533.3724V}, \cite{2024MNRAS.527.6110O}, \cite{Allen+2024} and \cite{yang2025cosmoswebunravelingevolutiongalaxy}.}
    \label{fig-size:BT}
\end{figure*}

In this section, we show the evolution of the size with redshift for galaxies classified by their morphology type (namely, bulge- and disk-dominated). We analyzed the radii from the two independent models: the effective radius $r_e$ from the Sérsic model, and the disk effective radius $r_{disk}$ and bulge effective radius $r_{bulge}$ from the B+D model. We computed size in kpc by converting the measured angular radius, using the \zphot\ given in the DJA \EAZY catalogs.

In Fig.~\ref{fig-size:BT}, we show the size evolution with \zphot\ for the Sérsic effective, disk, and bulge size measures for galaxies with $\log{M_\star/\si{\Msun}} > 10$, which corresponds to a mean mass of $\approx4\times10^{10} \, \si{\Msun}$. We show the trend for all galaxies, disk- and bulge-dominated as defined in Sect. \ref{sub:morpho-uvj}. We note that our sizes were computed as the average size over the $\sim1-5 \si{\um}$ range, whereas size-redshift evolution studies typically use the size measured in a given band that probes the same rest-frame optical range over the studied redshift range. This means that our measurements probe the rest-frame optical at $z\gtrsim0.5$; however, this approach may introduce differences when comparing our results with studies that base their size measurements on a single band for a defined redshift range.

Our measurements show that the overall size ($r_e$, from the Sérsic model) of all galaxies with $\log{M_\star/\si{\Msun}} > 10$ increases with redshift from about 1 kpc at $z\sim5$ to $\sim 2.5$ kpc at $z\sim0.5$. 
Disk-dominated galaxies show larger sizes by about $0.1-0.2$ kpc compared to the whole sample, and increase with time, while bulge-dominated galaxies are significantly smaller ($\sim1$ kpc) with little to no evolution out to $z\sim1.7$, but a relatively steep increase in size out to $\sim 2$ kpc at $z\sim0.6$.  

To quantify the size-redshift evolution, we fitted a ${r_e = R(1+z)^a}$ model, commonly used in the literature, to our measurements of the Sérsic effective radius. The fit was performed using the least squares method through the \texttt{lmfit} Python package, which returns the uncertainty of each parameter. Fitting the model to all galaxies yields $r_e = (3.66\pm0.04)(1+z)^{-0.59\pm0.01} \rm kpc$. By restricting the fit to the disk-dominated galaxies population, we obtain $r_e = (4.02\pm0.04)(1+z)^{-0.61\pm0.01} \rm kpc$. For the bulge-dominated population, the fit is $r_e = (3.34\pm0.10)(1+z)^{-0.87\pm0.04} \rm kpc$.

We compared our results with the literature based on \JWST\ measurements, namely, \cite{2024MNRAS.533.3724V}, who measured size evolution for Lyman break galaxies (LBGs) selected at $\log{M_\star/\si{\Msun}} > 9$, and \cite{2024MNRAS.527.6110O}, who measured the size evolution for $\log{M_\star/\si{\Msun}} > 9.5$ galaxies. We also compared with the results from \cite{Allen+2024} who measured the size evolution of $M_\star = 5\times10^{10}\, \si{\Msun}$ galaxies, which was obtained by scaling all galaxies to the same mass using the mass-size relationship. This comparison shows a good consistency with the literature. However, we note that, because of the strong dependence of galaxy size on stellar mass, the different mass limits and sample selection used in the literature complicate the comparison and are likely the reason for the differences.

The middle and right panels of Fig.~\ref{fig-size:BT} show the evolution of the disk and bulge sizes of galaxies from the two-component (B+D) model fit. This shows that the disk size increases from $\sim 1.2$ kpc to $\sim 3.4$ kpc from $z\sim5$ to $z\sim0.5$, with the disk-dominated population exhibiting larger disks by approximately 0.2 kpc, as expected. The bulge-dominated population may still host some disk components, which remain small ($\sim 1$ kpc) out to $z\sim 2.5$ and increase to $\sim 3.0$ kpc by $z\sim0.6$. This suggests that the disk components of bulge-dominated galaxies begin to grow their $r_{disk}$ at later times.
Bulges, on the other hand, show little-to-no growth with redshift with the total population exhibiting bulge sizes of about 1 kpc. However, the bulge component of the bulge-dominated population displays a mild increase in  $r_{bulge} of \sim0.8$ kpc from $z\sim 5$ to $z\sim 0.6$,  reaching $r_{bulge} \sim 1.6$ kpc. The associated $1\, \sigma$ dispersion is of comparable magnitude, making it difficult to draw robust conclusions.

\begin{figure*}[t!]
    \centering
    \includegraphics[width=0.99\textwidth]{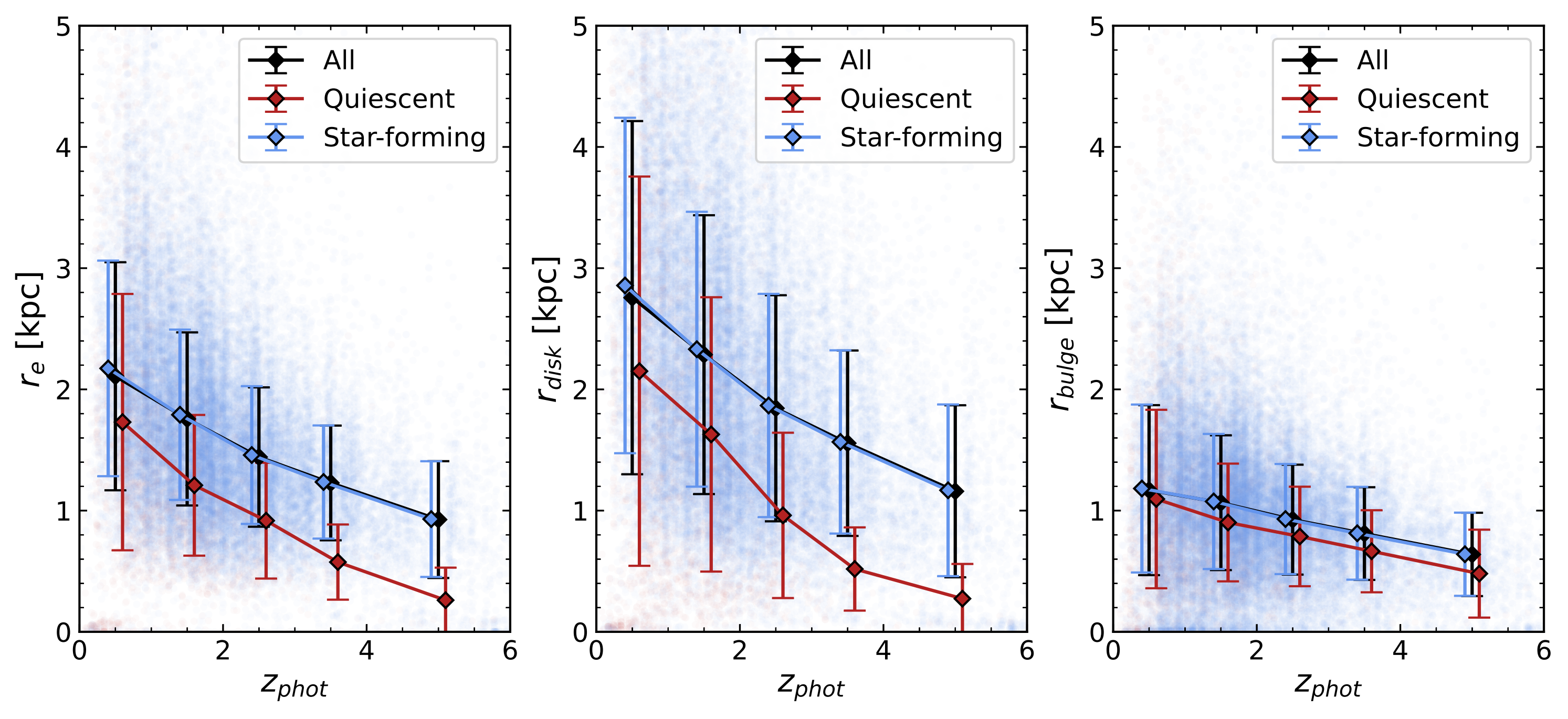}
    \caption{ Evolution of the sizes of quiescent and star forming galaxies with $\log{M_\star/\si{\Msun}} > 10$. The plots show the Sérsic effective radius $r_e$, the disk effective radius $r_{disk}$, and the bulge effective radius $r_{bulge}$ as a function of \zphot. Galaxies are classified as quiescent or star-forming based on their $UVJ$ colors. The points and error bars indicate the mean and $1\, \sigma$ dispersion in the corresponding \zphot\ bin.}
    \label{fig-size:quiescence}
\end{figure*}

We also investigated the size evolution of quiescent and star-forming galaxies selected using their $UVJ$ colors in Fig.~\ref{fig-size:quiescence}. Overall, as expected, quiescent galaxies have smaller sizes than star-forming galaxies. The sizes of quiescent galaxies increase with redshift, from $\sim 0.2$ kpc for the earliest quiescent galaxies at $z\sim5$ to $\sim 2$ kpc by $z\sim 0.6$. Star-forming galaxies are larger and increase in size from $\sim 1.2$ kpc to $\sim 2.6$ kpc by $z\sim0.5$. This shows that the size growth of quiescent galaxies is steeper than that of star-forming galaxies.  This size evolution is consistent with previous studies showing that both populations experience growth over time, driven by minor mergers and continuous star formation in star-forming galaxies, whereas quiescent galaxies primarily grow via dry mergers \citep[e.g.,][]{2014ApJ...788...28V, Whitaker2017}. While high-redshift quiescent galaxies are compact at formation, often referred to as "red nuggets" \citep[e.g.,][]{Damjanov2009}, Fig.~\ref{fig-size:quiescence} indicates that they continue to grow in size down to $z\sim0.5$, a trend that can be driven by minor mergers and other accretion processes \citep[e.g.,][]{Newman2012, Belli2015}.
However, part of the observed size growth of quiescent galaxies may be attributed to progenitor bias -- as new, larger galaxies continue to quench and enter the quiescent population at later times, they increase the average size of the population, making the size evolution appear more dramatic than if individual galaxies were tracked over time \citep[e.g.,][]{Barro2013, Carollo2013}.

The disk sizes of star-forming galaxies increase from $\sim1.4$ kpc at $z\sim5$ to $3.4$ kpc by $z\sim0.5$, consistent with expectations from inside-out growth models, where gas accretion and star formation preferentially occur in the outskirts \citep[e.g.,][]{2012ApJ...748L..27P, Morishita2014, Matharu2024}. The disk sizes of quiescent galaxies remain small and are similar to, or smaller than, their bulges, at least out to $z\sim2$, where the disk size increases to about $2.6$ kpc. However, the scatter is significantly higher, making it difficult to draw meaningful conclusions about the disk sizes of quiescent galaxies. The bulge sizes for both star-forming and quiescent galaxies exhibit a mild evolution with redshift. Star-forming galaxies grow by about $0.6$ kpc, reaching $\sim 1.2$ kpc by $z\sim0.5$. Quiescent galaxies are slightly more compact and, at high redshift, have bulges of about $0.4$ kpc, increasing to about $1.2$ kpc by $z\sim0.6$. The persistence of small bulge sizes in both populations suggests that bulges reach their final configuration early, while the surrounding disks continue to evolve, particularly in star-forming galaxies. This is consistent with models in which early compaction events, such as mergers or disk instabilities, form a central bulge, after which the fate of the galaxy depends on the availability of fresh gas for continued star formation or quenching mechanisms \citep[e.g.,][]{Barro2017, Tacchella2018}.

\subsection{Evolution of the stellar mass surface density for quiescent and star-forming galaxies}

\begin{figure*}[t!]
    \centering
    \includegraphics[width=\textwidth]{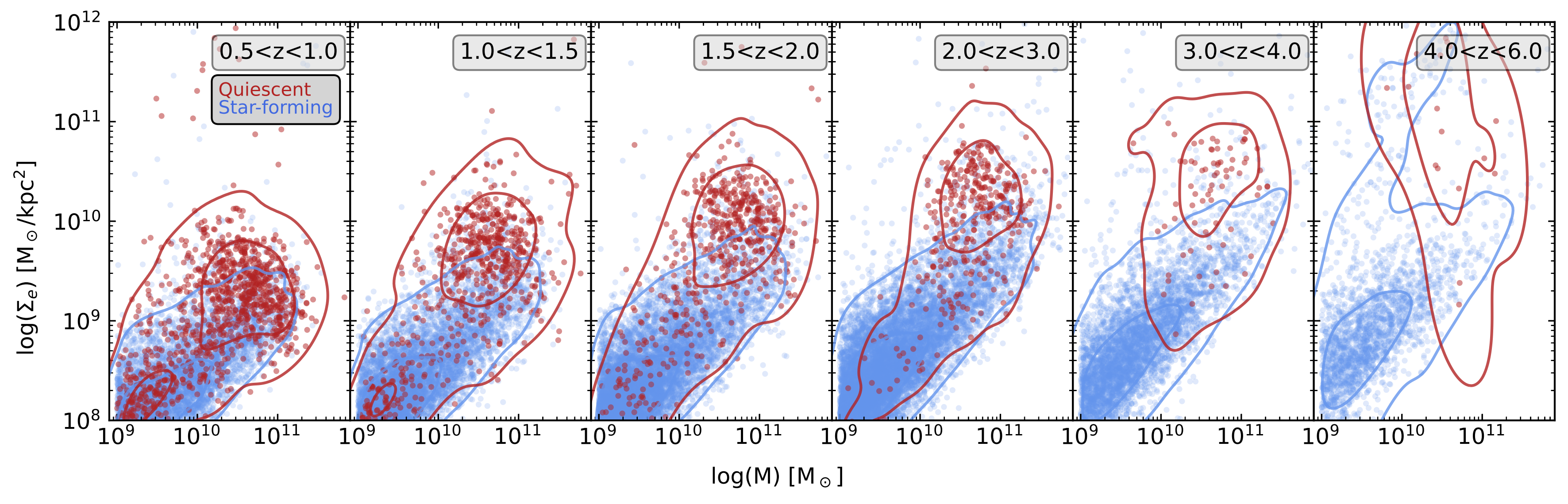}
    \caption{ Evolution of the stellar mass surface density, $\Sigma_e$. Shown are masses and redshifts from the DJA, calculated using \EAZY, and effective radii, $r_e$, measured by \sepp in this work. Galaxies are classified as quiescent or star-forming according to their position in the UVJ diagram. The contour lines result from kernel density estimation using \texttt{seaborn.kdeplot}. The data for this figure includes 45,201 galaxies such that $\log{M_\star/\si{\Msun}} > 9$.}
    \label{fig:surface-density}
\end{figure*}

To investigate the relationship between the compactness of a galaxy and its star formation activity, we calculated the stellar mass surface density, $\Sigma_e$, which is the mass contained within the Sérsic effective radius, given by $\Sigma_e = M_{\star} / 2 \pi r_e^2$. Fig.~\ref{fig:surface-density} shows $\Sigma_e$ as a function of stellar mass for quiescent and star-forming galaxies of $\log{M_{\star}/\si{\Msun}}>9$, to facilitate comparison with the existing literature, in six redshift bins at $0.5<z<6$. This figure demonstrates that quiescent galaxies are more compact, with higher surface mass densities than those of star-forming galaxies, and that quiescent galaxies become increasingly dense at earlier times. This finding is in good agreement with previous work, e.g., \cite{Barro2017}, which shows this relation out to $z\sim3$. Our results indicate that this relation extends out to $z\sim5$ and that quiescent galaxies increase in compactness the earlier they form. This is consistent with the observed compactness of some of the earliest quiescent galaxies found by \JWST \citep{Carnall2023, deGraaff2024, Weibel2024b, Ito2024, Wright2024}.

\section{Conclusions}\label{sec:conclusions}

This work presents a catalog of galaxy morphologies measured from \JWST imaging of the major extragalactic surveys CEERS, GOODS, and PRIMER, processed through the DAWN JWST Archive. The catalog contains morphology information for more than 340,000 sources, detected on NIRCam LW stacks, with photometric redshifts extending to $z>10$. We fitted each source with two independent models of brightness profiles: a single Sérsic model and a Bulge+Disk model, using \sepp. 

To validate our measurements, we compared our results with those from the literature obtained using independent methods and software, and we find good consistency. To demonstrate the scientific application, we used our morphological measurements in combination with the photo-z and physical parameters from the DJA \EAZY catalogs. We investigated the relation between galaxy morphology, $UVJ$ colors as an indicator of star formation activity, and redshift. We summarize our results as follows.
\begin{itemize}

\item Using $UVJ$ color selection, we find that early-type, bulge-dominated galaxies predominantly reside in the quiescent region, whereas late-type, disk-dominated galaxies exhibit blue, star-forming colors. We also observe a bimodality in these distributions, with some populations of quiescent galaxies that are disk-dominated, as well as star-forming galaxies that are bulge-dominated.

\item The two-component fits reveal that low $B/T$ galaxies preferentially occupy the star-forming $UVJ$ region, while high $B/T$ galaxies populate the quiescent region. At $z>3$, however, we observe a population of high $B/T$ and bulge-dominated galaxies, consistent with a blue nugget phase.

\item The Sérsic effective radius ($r_e$), disk effective radius ($r_{\rm disk}$), and bulge effective radius ($r_{\rm bulge}$) all show a decreasing trend with increasing redshift. Star-forming galaxies exhibit systematically larger sizes compared to quiescent galaxies at all redshifts, consistent with prior studies. Quiescent galaxies, while smaller than star-forming ones, show a steeper increase in their effective radius with time.

\item Quiescent galaxies are significantly more compact than their star-forming counterparts, leading to high stellar mass surface densities ($\Sigma_e$). We find that $\Sigma_e$ for quiescent galaxies is nearly an order of magnitude higher at $z \sim 4$ compared to $z \sim 1$, consistent with the observed compactness of some of the earliest quiescent galaxies observed by \JWST.

\end{itemize}

This morphological catalog is a valuable addition to the DJA, enabling a range of in-depth studies of the morphological transformations associated with galaxy evolution.

\section*{Data availability}

Our catalog is available in electronic form at the CDS via anonymous ftp to cdsarc.u-strasbg.fr (130.79.128.5) or via \href{http://cdsweb.u-strasbg.fr/cgi-bin/qcat?J/A+A/699/A343}{http://cdsweb.u-strasbg.fr/cgi-bin/qcat?J/A+A/699/A343}. It is also available from the \href{https://dawn-cph.github.io/dja/blog/2024/08/16/morphological-data/}{DAWN JWST Archive}. The code used to run the model fitting and the notebooks used to analyze the results are available on \href{https://github.com/AstroAure/DJA-SEpp}{GitHub} under the \href{https://www.gnu.org/licenses/gpl-3.0.html}{GNU GPL v3.0 license}.

\begin{acknowledgements}

The data products presented herein were retrieved from the DAWN JWST Archive (DJA). DJA is an initiative of the Cosmic Dawn Center (DAWN), which is funded by the Danish National Research Foundation under grant DNRF140.
This work has received funding from the Swiss State Secretariat for Education, Research and Innovation (SERI) under contract number MB22.00072
This work has been conducted during the research internship of Aurélien Genin at the Cosmic DAWN Center, under the supervision of Marko Shuntov.
This internship has also been made possible thanks to the financial support of the "Space: Science and Challenges of Space" chair at Ecole polytechnique, financed by ArianeGroup and Thales Alenia Space, as well as an Erasmus+ internship grant.

\end{acknowledgements}

\bibliographystyle{aa}
\bibliography{draft}

\appendix

\section{PSF results}\label{sec:appendix-psf}

\begin{figure}[!ht]
    \centering
    \resizebox{\hsize}{!}{\includegraphics{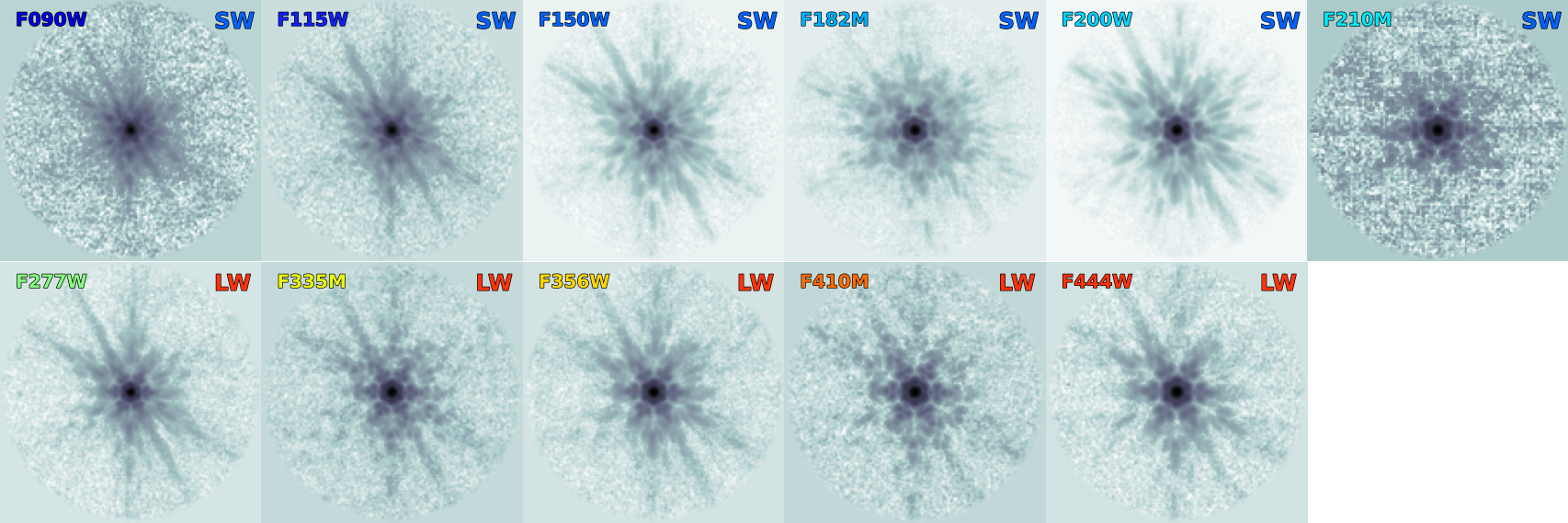}}
    \caption{\footnotesize PSFs calculated with \psfex for different bands of \JWST on the GOODS-S field. The point-like sources used are selected using the starline method. The name of the band (resp. channel) is written in the top right (resp. left) of each image. The images are displayed in logarithmic scale and color-inverted.}
    \label{fig-psf:psf}
\end{figure}

In Fig.\ref{fig-psf:psf}, we show the point spread functions (PSF) calculated over the GOODS-S field in different bands. As presented in Sect. \ref{sub:psf}, these empirical PSFs have been generated using \psfex and a point-like sources selection based on \texttt{MU\_MAX} and \texttt{MAG\_AUTO} measurements.

\section{\sepp model priors}\label{sec:appendix-sepp-priors}

\begin{figure}[!ht]
\begin{center}
\includegraphics[width=0.95\columnwidth]{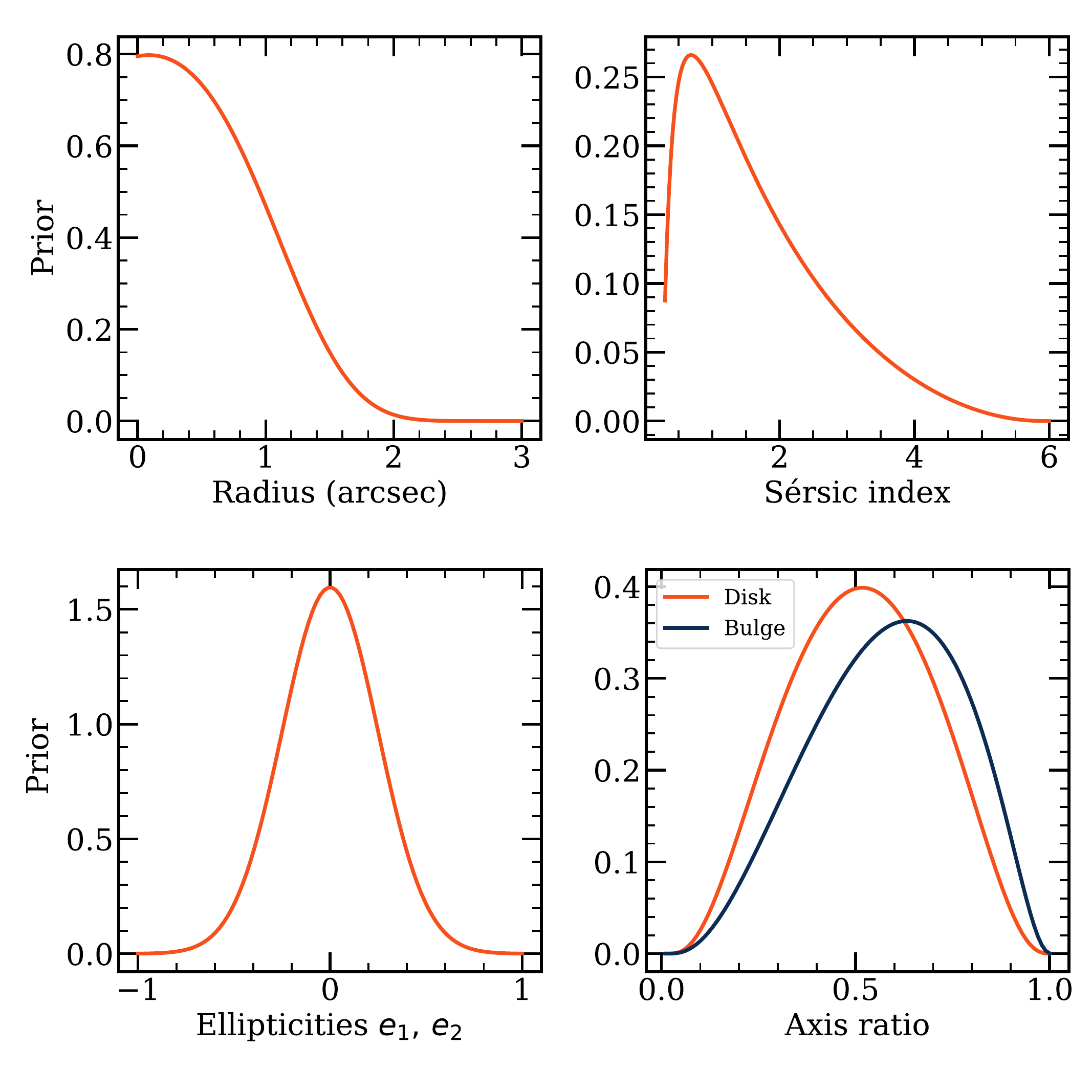}
\caption{Priors that we adopt for the source model parameters.}
\label{fig:priors}
\end{center}
\end{figure}

In Fig.~\ref{fig:priors}, we show the priors used in our Sérsic and Bulge+Disk models (Sect. \ref{sub:models}) for \sepp. The first three panels correspond to the priors on the effective radius (for both models), Sérsic index and ellipticities for the Sérsic model, while the fourth panel shows the priors on the axis ratio that we adopt for the B+D fits.

\newpage
\section{Usage of Amazon Web Services}\label{sec:appendix-AWS}

This work made use of Amazon Web Services (AWS), in particular the \href{https://aws.amazon.com/ec2/}{EC2}\footnote{Elastic Compute Cloud} and \href{https://aws.amazon.com/s3/}{S3}\footnote{Simple Storage Service} services. AWS EC2 offers on-demand cloud computing, with the possibility to choose very different machines (refers as instances) with various resources in terms of computing power or memory. AWS S3 is a cloud storage service, already used to store the DJA. Here, we provide more details on the computing and use of AWS.

\subsection{AWS EC2, VS Code, Jupyter interfacing}\label{sub-AWS:VSJupytEC2}

For simplicity of use, we developed some basic bash scripts to allow for easy interfacing between AWS EC2, VS Code and Jupyter. These scripts enable to start an EC2 instance, launch a Jupyter server in it, and connect it to VS Code. The user can then use VS Code seamlessly to code, while running Jupyter notebooks directly on the EC2 instance. This has the added benefit of keeping the notebooks running, even if VS Code and the user computers are shutdown, which proves very useful for \sepp as it takes many hours to run. This code has been made available under the \href{https://www.gnu.org/licenses/gpl-3.0.html}{GNU GPL v3.0 license} on GitHub\footnote{\url{https://github.com/AstroAure/VSJupytEC2}}.

\subsection{Automation}\label{sub-AWS:automation}

This section details the different steps used in this work to generate a morphological measurements catalog on one field. Step 1 is the pre-processing with PSF estimation and tiling. Step 2 is the result of the tiling process to run \sepp in parallel. Steps 3 and 4 are the merging of catalogs between all the tiles, the Sérsic and B+D models and the DJA. These steps are detailed in the following process:

\begin{enumerate}
    \item Main instance (\textit{m5d.4xlarge} for memory)
    \begin{enumerate}
        \item Download all the frames of the field from the DJA;
        \item Pre-process them (decompress and save to a S3 bucket);
        \item Find point-like sources, estimate the PSFs in all the bands and save them to a S3 bucket;
        \item Tile the images and save them to a S3 bucket.
    \end{enumerate}
    \item For every tile
    \begin{enumerate}
        \item Start an instance (by default, \textit{c6a.4xlarge});
        \item Download the necessary files (frames, PSFs, configuration files, code);
        \item Run \sepp with the selected model (Sérsic or Bulge+Disk);
        \item Save the resulting catalog and images to a S3 bucket.
    \end{enumerate}
    \item Main instance (\textit{m5d.4xlarge} for memory)
    \begin{enumerate}
        \item Merge tile catalogs and images;
        \item Save the full catalog and images to a S3 bucket.
    \end{enumerate}
    \item Main instance (\textit{m5d.4xlarge} for memory)
    \begin{enumerate}
        \item Merge the full \sepp catalogs with the two models (Sérsic and Bulge+Disk) to the DJA catalogs;
        \item Save the complete catalog to the DJA.
    \end{enumerate}
\end{enumerate}

Each of the sub-steps in this process are coded as bash and/or Python scripts. The \texttt{dja\_sepp} Python package has also been developed to help with many of the different operations involved. It is available through \texttt{pip} and on GitHub\footnote{\url{https://github.com/AstroAure/DJA-SEpp}}. This repository also holds many Jupyter notebooks used for the pre- and post-processing steps, as well as for the analysis of results (Sect. \ref{sec:results}).

\end{document}